\documentclass[onecolumn,superscriptaddress,showpacs,pra]{revtex4}
\usepackage{epsfig}
\usepackage{epstopdf}
\usepackage{delarray}
\usepackage{amsmath, amssymb}
\usepackage{bm}
\usepackage{graphicx}
\usepackage{color} 
\usepackage{amssymb}

\usepackage[figuresright]{rotating}
\usepackage{float} 
\usepackage{lipsum}
\usepackage{epstopdf} 
\usepackage{amsmath}
\usepackage[section]{placeins}

\begin{document}



\title[Efficient Sensing of the von Kármán Vortices Using Compressive Sensing]{Efficient Sensing of the von Kármán Vortices Using Compressive Sensing}

\author{Cihan Bay\i nd\i r}
\affiliation{\.{I}stanbul Technical University, Engineering Faculty, 34469 Maslak, \.{I}stanbul, Turkey.\\  
Bo\u{g}azi\c{c}i University, Engineering Faculty, 34342 Bebek, \.{I}stanbul, Turkey. \\
CERN, CH-1211 Geneva 23, Switzerland.}
\email{cbayindir@itu.edu.tr}

\author{Bar\i\c{s} Naml\i}
\affiliation{\.{I}stanbul Technical University, Engineering Faculty, 34469 Maslak, \.{I}stanbul, Turkey.}
\email{barisnamli17@gmail.com}

\begin{abstract}
In this paper, we discuss the usage and implementation of the compressive sensing (CS) for the efficient measurement and analysis of the von Kármán vortices. We consider two different flow fields, the flow fields around a circle and an ellipse. We solve the governing $k-\epsilon$ transport equations numerically in order to model the flow fields around these bodies. Using the time series of the drag, $C_D$, and the lift, $C_L$, coefficients, and their Fourier spectra, we show that compressive sampling can be effectively used to measure and analyze Von Kármán vortices. We discuss the effects of the number of samples on reconstruction and the benefits of using compressive sampling over the classical Shannon sampling in the flow measurement and analysis where Von Kármán vortices are present. We comment on our findings and indicate their possible usage areas and extensions. Our results can find many important applications including but are not limited to measure, control, and analyze vibrations around coastal and offshore structures, bridges, aerodynamics, and Bose-Einstein condensation, just to name a few.
\end{abstract}

\maketitle


\section{Introduction}
Vortex street generated in the wake region of flow around a bluff body is known as von Kármán vortices. These vortices impose periodic pressure variations on the bluff body in the transverse and parallel to the flow directions. These pressure variations induce vibrations on the bluff body in the flow, having an excitation force spectra centered at the von Kármán vortex shedding frequency. Undoubtedly, von Kármán vortices is one of the most widely studied and well-known phenomena in the field of fluid dynamics, and literature on this subject is vast \cite{Vonkarmanbook, MunsonYoungOkiishi, KunduCohen, Zdravkovich97, Zdravkovich02}. When engineering structures are buffeted by steady winds and waves, they experience pushes and pulls in the direction perpendicular to the flow due to these vortices. Besides the well-known simple bluff bodies such as rectangular, circular, and elliptical cylinders, the von Kármán vortices in the flow field around many different engineering structures are studied in the literature. These structures include but are not limited to offshore structures \cite{Chella2012}, tall buildings \cite{Simiu1986}, bridges \cite{Scanlan}, wind turbines \cite{Lynum}, airplanes and space shuttles \cite{Anderson2010}, underwater robots and artificial fish \cite{Toming}, just to name a few. von Kármán vortices are also observed in nature, for example in the air flows around islands \cite{Mizota} and around an impacting drop \cite{Thoraval}. It is possible that different types of hydrodynamic media other than air and water can exhibit von Kármán vortex type structures. For example, they are observed in atomic superfluid gas and around the obstacles in atomic Bose-Einstein condensates \cite{SasakiSuzuki, KwonKim, Stagg}. These vortices can also be observed in acoustic \cite{KimDurbin} as well as the electrically and magnetically mediated media \cite{Dutta}, just to name a few. 

Generally, von Kármán vortices can impose undesired vibration on the engineering structures and may eventually lead to resonance and fatigue. Therefore, various methodologies have also been proposed to mitigate the von Kármán vortices. Wu et. al. presented a suppression technique for the von Kármán vortex street behind a circular cylinder using a traveling wave generated by a flexible surface \cite{Wu2007}. Zhao et. al. used a traveling wave of a flexible wall to reduce the turbulent drag \cite{Zhao2004}. Patnaik and Wei proposed an angular momentum injection approach to tame and control the wake turbulence \cite{PatnaikPRL}. Ledda et. al. studied the suppression of von Kármán vortex streets past porous rectangular cylinders \cite{Ledda}. Although these vortices are generally not desired, the possibility of using them for piezoelectric energy harvesting is also considered \cite{Demoria}. While there are many different approaches to model the von Kármán vortices analytically and experimentally under various scenarios \cite{Ohle, Crowdy, Krishnan, Sakamoto}, due to complexity of the problem and the geometries researchers mainly use numerical approaches \cite{Sobey}.
This brief summary on von Kármán vortices is by no means complete, thus the reader is referred to the aforementioned papers and the references therein to catch a glimpse of the vastness of the subject.

On the other hand, the compressive sensing (CS) technique has emerged as one of the most successful theories in the signal measurement and analysis \cite{candes, candes2006compressive, candesRomberg2, candesTao, Baraniuk}. CS outperforms the classical sampling by using far fewer measurement for the exact reconstruction of a sparse signal. This technique allowed the development of efficient single-pixel cameras, analog-to-digital converters, seismographs, and widely used in sensor networks. Although the CS has revolutionized the field of signal processing, its implications in fluid dynamics are quite limited. Some of the work done in this context are summarized below. Bright et. al. proposed a CS-based machine learning strategy to characterize flow fields around a cylinder using limited pressure data in \cite{BrightLin}. A CS-based tomographic sensing approach for experimental fluid dynamics studies is introduced by Petra and Schnörr in \cite{Petra}. Brunton et. al. discussed the sensing and dynamic mode decomposition of a chaotic mixing model, the double gyre flow field, by CS in \cite{Brunton}. Bai et. al. introduced a low-dimensional approach for reconstruction of airfoil data using CS \cite{Bai2}. Kramer et. al. used CS and dynamic mode decomposition-based techniques for the identification of flow regimes and bifurcations in complex flow fields \cite{Kramer}. The possible usage of CS for the efficient computational analysis in nonlinear wave simulations is discussed by one of us in \cite{Baysci}. Measurement and early detection of 1D and 2D rogue waves observed in hydrodynamic media are also discussed by one of us in \cite{BayEarlyCS} and \cite{BayTomog}, respectively. Later, Malara et. al. analyzed the usage of the CS for the extrapolation of random waves \cite{Malara}.

In this paper, we investigate the implementation and usage of the CS, its benefits, and possible application areas for the measurement and analysis of the von Kármán vortices. With this motivation, we consider two different flow fields around two different bluff bodies, a circular cylinder and an elliptical cylinder. We numerically simulate the flow fields around these bluff bodies and produce the time series of the drag coefficient, $C_D$, and the lift coefficient, $C_L$. Since these time series have a sparse representation in the Fourier domain, we show that CS can be efficiently used to construct these time series using a far fewer number of measurements than the classical Shannon states. We discuss the possible advantages of the CS for the measurement and analysis of similar time series of various flow parameters. We also discuss potential applications of CS for analyzing fluid flows exhibiting von Kármán vortices and similar phenomena.

\section{Methodology}

\subsection{Numerical Approach for the Generation of Von-Karman Vortices}
In order to model the time series of the drag and the lift coefficients of the flow around the circular and elliptical blunt bodies, we use the ANSYS Fluent software and use the $k-\epsilon$ turbulence model \cite{ANSYSFluent, Launder}. Here,  $k$ is the turbulent kinetic energy and $\epsilon$ is the rate of dissipation and this model is based on two model transport equations for the evaluation of velocities $u_i$ in the form of
\begin{equation}
\frac{\partial (\rho k)}{\partial t}+ \frac{\partial (\rho k u_i)}{\partial x_i}=\frac{\partial }{\partial x_j} \left[ \left(\mu+\frac{\mu_t}{\sigma_k} \right) \frac{\partial k}{\partial x_j}  \right]+ G_k+G_b-\rho \epsilon -Y_M+S_k
\label{eq1}
\end{equation}

\begin{equation}
\frac{\partial (\rho \epsilon)}{\partial t}+ \frac{\partial (\rho \epsilon u_i)}{\partial x_i}=\frac{\partial }{\partial x_j} \left[ \left(\mu+\frac{\mu_t}{\sigma_{\epsilon}} \right) \frac{\partial \epsilon}{\partial x_j}  \right]+ C_{1\epsilon}\frac{\epsilon}{k}(G_k+C_{3\epsilon}G_b)-G_{2\epsilon} \rho \frac{\epsilon^2}{k}+S_{\epsilon}
\label{eq2}
\end{equation}
Here, $\rho$ is the density, $k$ is the turbulent kinetic energy and $\epsilon$ is the rate of dissipation. The turbulent (eddy) viscosity is defined as $\mu_t=\rho C_{\mu} k^2/ \epsilon$. In Eq.(\ref{eq1}) and Eq.(\ref{eq2}), generation of the turbulence kinetic energy due to mean velocity gradients is represented by the term $G_k=-\rho \frac{\partial u_j}{\partial x_i} \overline{u'_i u'_j}$ and the generation of the turbulence due to buoyancy is represented by the term $G_b=\beta g_i \frac{\mu_t}{Pr_t} \frac{\partial T}{\partial x_i}$ where $\beta$ shows the thermal expansion constant, $Pr_t$ is the turbulent Prandtl number for energy, $T$ is temperature and $g_i$ is the component of the gravitational acceleration in the $i^{th}$ direction \cite{ANSYSFluent}. The parameter $C_{3\epsilon}=\tanh{|v/u|}$, where $v$ and $u$ show the flow velocities parallel and perpendicular to the gravity, shows the degree to which the rate of dissipation is effected by buoyancy. The dilatation dissipation rate is $Y_M=2\rho \epsilon M^2_t$ where $M_t=k^{0.5}/c$ is the turbulent Mach number \cite{ANSYSFluent}. The default constants implemented in the ANSYS Fluent software are $C_{1\epsilon}=1.44, C_{2\epsilon}=1.92, C_{\mu}=0.09$ and the Prandtl numbers are $\sigma_k=1.0$ and $\sigma_{\epsilon}=1.3$. In these formulations, $S_k$ and $S_{\epsilon}$ are source terms, which are selected as $0$ for this study. ANSYS Fluent solves these equations using the finite volume method. The reader is referred to the \cite{ANSYSFluent, Launder} for a more comprehensive discussion of the $k-\epsilon$ turbulence model and to \cite{ANSYSFluent} for a detailed discussion of the implementation of the ANSYS Fluent. In this paper, we simulate the 2D horizontal flow of air around two types of blunt bodies, namely a circular cylinder and an elliptical cylinder by solving the transport equations summarized above. Parameter selection is discussed in the coming sections of this paper.

\subsection{Review of the Compressive Sampling}
\noindent After its introduction to the field of mathematical signal processing less than two decades ago, compressive sensing (CS) proved to be a revolutionary algorithm \cite{candes, candes2006compressive, candesRomberg2, candesTao, Baraniuk}. Currently, it is commonly used as one of the most successful algorithms of signal and image processing. Although the majority of the literature on the applications of the CS are theoretical, many possible technologies have already been developed including but are not limited to accurate and efficient sensing and analysis devices such as the single-pixel cameras, analog-to-digital converters, seismographs, and CS-based tomography devices, just to name a few. In this section of the paper, we try to give a brief summary of the implementation of the CS algorithm \cite{candes}.

Let the function $\eta$ represent a $K$-sparse signal with $N$ entries. That is, only $K$ out of $N$ entries of the signal are nonzero. By utilizing an orthogonal transformation matrix ${\bf \Psi}$, it is possible to represent $\eta$ in a transformed domain in terms of the orthogonal basis functions. Commonly used orthogonal transformation to obtain such a representation are the Fourier, wavelet, or DCTs, and each can have pros and cons depending on providing a sparse representation of the signal. Using these and similar transformations, one can represent the signal as $\eta= {\bf \Psi} \widehat{ \eta}$, where $\widehat{ \eta}$ is a coefficient vector. After discarding the zero entries of the signal $\eta$, it is possible to get $\eta_s= {\bf \Psi}\widehat{ \eta}_s$, where the function $\eta_s$ includes non-zero components of the signal only.

The CS theory proves that any $K$-sparse with $N$ entries can be reconstructed with an overwhelmingly high probability using $M \geq C \mu^2 ({\bf \Phi},{\bf \Psi}) K \textnormal{ log (N)}$ random measurements  \cite{candes}. Here, the parameter $C$ shows a positive constant. The function $\mu^2 (\Phi,\Psi)$ indicates the mutual coherence between the transformed basis ${\bf \Psi}$ and sensing basis ${\bf \Phi}$ \cite{candes}. Therefore, by taking $M$ random projections and using the sensing matrix of ${\bf \Phi}$, it is possible to reconstruct $g={\bf \Phi} \eta$. In other words, the CS reconstruction problem can be restated as
\begin{equation}
\textnormal{ min} \left\| \widehat{ \eta} \right\|_{l_1}   \ \ \ \  \textnormal{under constraint}  \ \ \ \ g={\bf \Phi} {\bf \Psi} \widehat{ \eta}
\label{eq10}
\end{equation}
with $\left\| \widehat{ \eta} \right\|_{l_1}=\sum_i \left| \widehat{ \eta}_i\right|$. Therefore, among all the possible signals which satisfy the given constraints, the ${l_1}$ solution of the CS problem is  $\eta_{{}_{CS}} ={\bf \Psi} \widehat{ \eta}$. 

In the literature, there are also other optimization techniques such as greedy pursuit or re-weighted $l_1$ minimization algorithms which can be used for the reconstruction of a sparse signal \cite{candes, candes2006compressive, candesRomberg2, candesTao, Baraniuk}. For a more comprehensive discussion of the CS algorithm the reader is referred to \cite{candes, candes2006compressive, candesRomberg2, candesTao, Baraniuk} and the references therein. 

An investigation of the existing vast data about the von Kármán vortices and associated parameters suggests that time series of these vortices and similar phenomena generally contain few fundamentals harmonics, thus it is possible to obtain their sparse representation in Fourier, DCT or wavelet space. Therefore, the CS algorithm can be efficiently employed to reconstruct these data using a far fewer number of measurements ($M$) than the classical Shannon's theorem states ($N$). In this paper, we numerically generate the time series of the drag and the lift coefficients, $C_D$ and $C_L$, around two different types of blunt bodies, namely a circular cylinder and an elliptical cylinder. Employing a spectral analysis using FFTs, we show that these time series have a sparse representation in the spectral domain. Thus, by taking $M$ random measurements in the time domain and applying the $l_1$ minimization, the CS can be used for the reconstruction of these types of time series. We present and discuss our findings in the next section of this paper.

\section{Results and Discussion}

\subsection{Results for the Flow Field Around A Circle}
\begin{figure}[h!]
\begin{center}
   \includegraphics[width=5.0in]{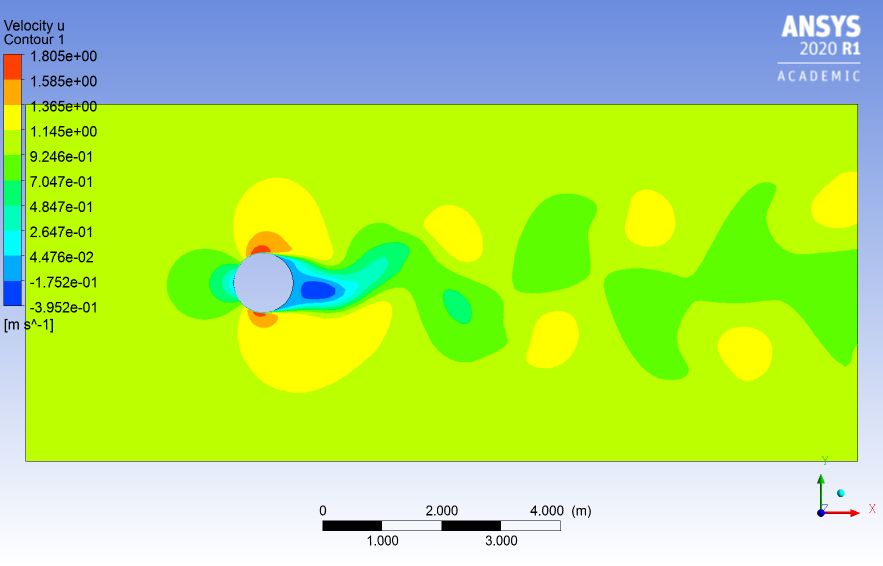}
  \end{center}
\caption{\small  von Kármán vortices around a circle for $v=1m/s$ at $t=200s$. }
  \label{fig1}
\end{figure}

We begin our analysis by considering a circle with a diameter of $D=1m$ as displayed in Fig.~\ref{fig1}. The upstream flow velocity for this simulation is taken to be $U=1m/s$. Density of air is selected as $\rho=1.225 kg/m^3$ and the viscosity of air is selected as $\mu=1.7894 \times 10^{-5} kg/(m.s)$. The corresponding Reynolds number for this simulation is $Re=\rho U D/ \mu \approx 6.85 \times 10^{4} $. In our simulations, we use a time step of $dt=0.1s$ and evolve the numerical simulations until time of $t=200s$. The turbulence model is selected to be the $k-\epsilon$ model.  The flow field around the circle exhibiting the von Kármán vortices is displayed in Fig.~\ref{fig1}.

\begin{figure}[h!]
\begin{center}
   \includegraphics[width=6.0in]{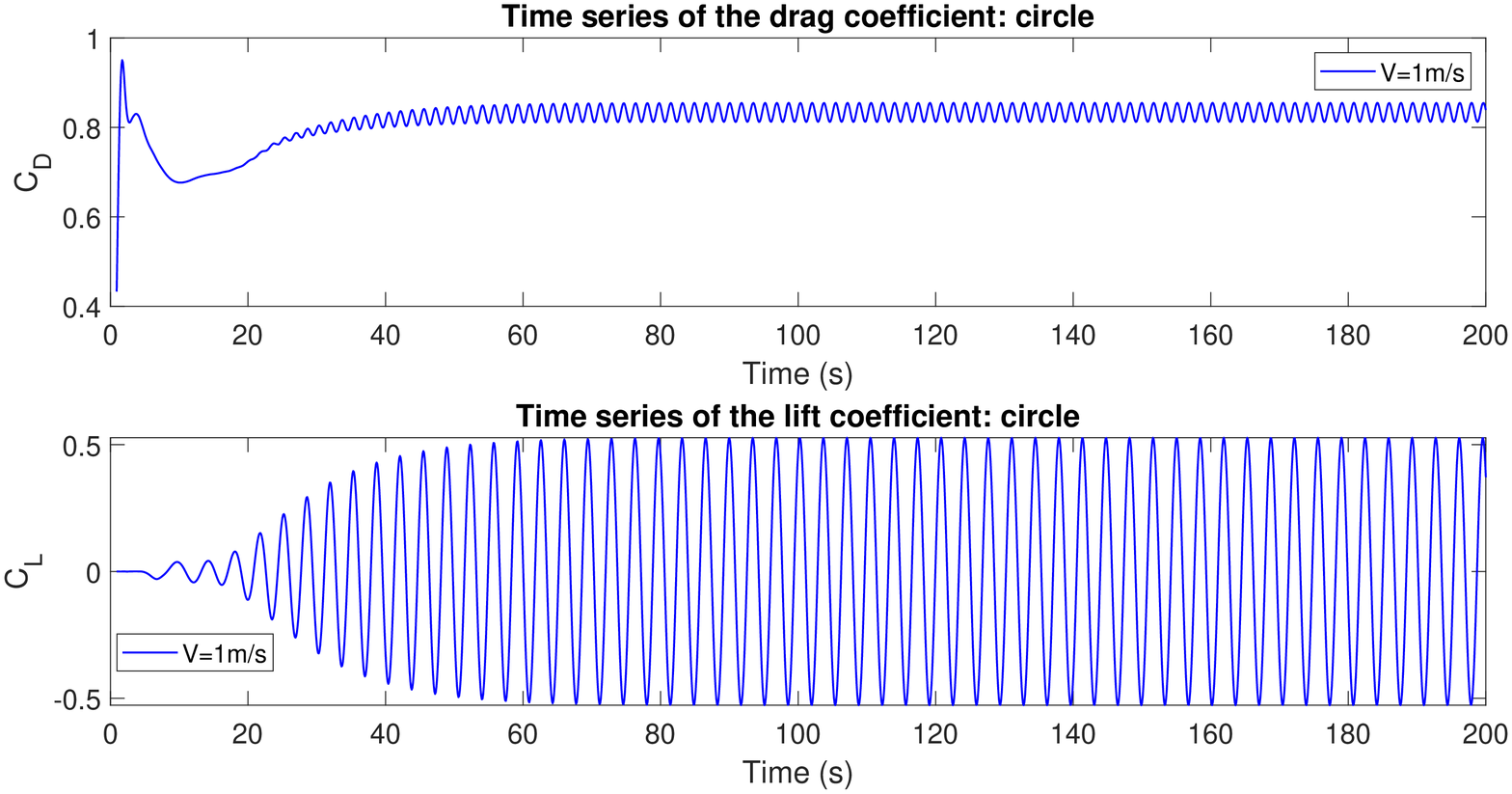}
  \end{center}
\caption{\small  Time histories of the von Kármán vortices around a circle for $v=1m/s$ a) Drag coefficient b) Lift coefficient.}
  \label{fig2}
\end{figure}

The time series of the drag and lift coefficients, namely $C_D$ and $C_L$, for this simulation is depicted in Fig.~\ref{fig2}. Checking this figure, it is possible to realize that the von Kármán vortices reach a steady-in-the-mean state after some adjustment time. The mean values of the steady-in-the-mean coefficients are $C_D \approx 0.835$ and $C_L \approx 0.000$, and their maximum values are $C_D \approx 0.852$ and $C_L \approx 0.505$. These results are in accordance with the well-established results tabulated for various values of Reynolds number \cite{MunsonYoungOkiishi, KunduCohen}. As expected, due to the existence of the von Kármán vortices in the flow field, the oscillations in the time series of $C_D$ and $C_L$ can be clearly observed confirming a dynamic behavior. 

\begin{figure}[h!]
\begin{center}
   \includegraphics[width=6.0in]{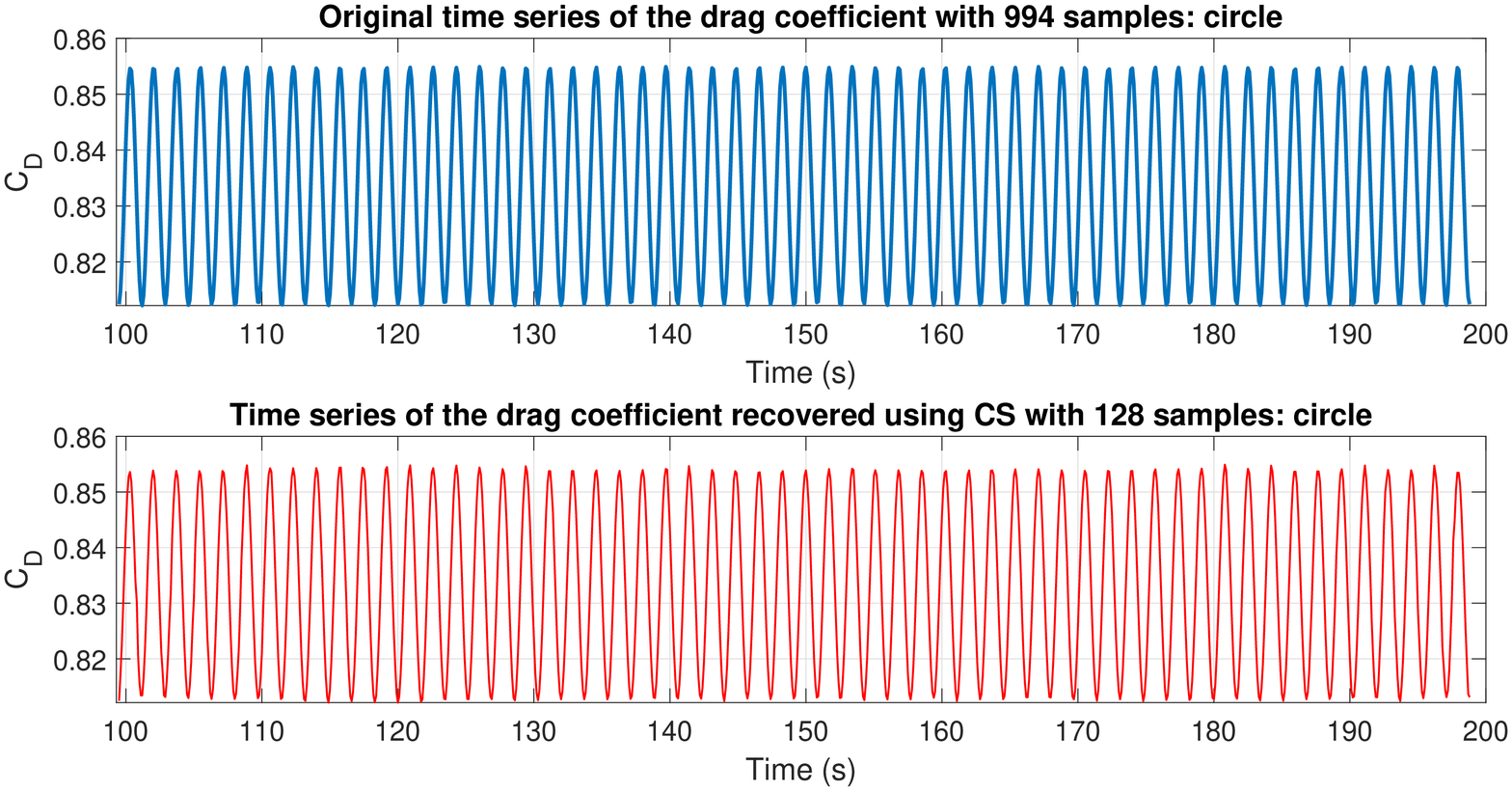}
  \end{center}
\caption{\small Spectrum of the time series of the drag coefficient, $C_D$ of the flow around the circle: a)  classically sampled using $N=994$ samples b) reconstructed using compressive sampling using $M=128$ samples.}
  \label{fig3}
\end{figure}

\begin{figure}[h!]
\begin{center}
   \includegraphics[width=6.0in]{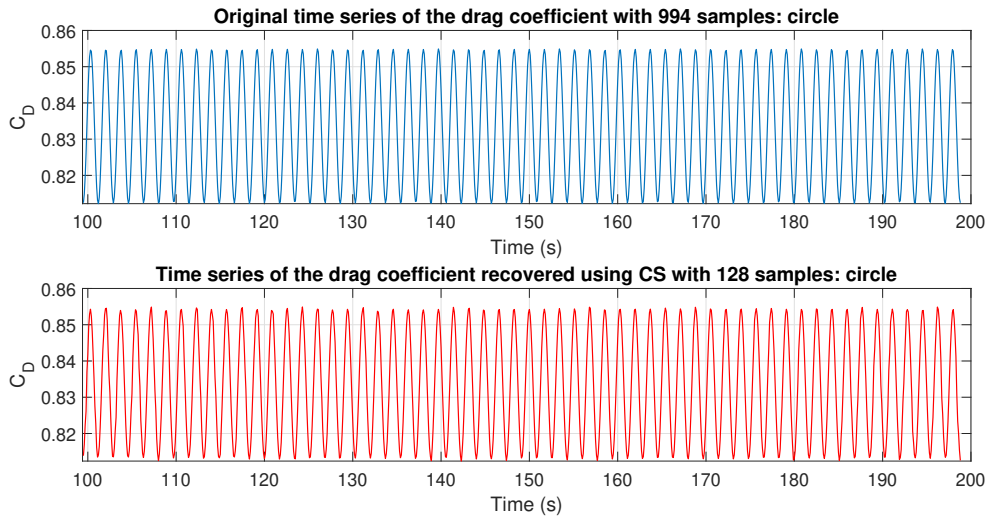}
  \end{center}
\caption{\small  Time series of the drag coefficient, $C_D$, of the flow around the circle: a) classically sampled using $N=994$ b) reconstructed using $M=128$ compressive samples.}
  \label{fig4}
\end{figure}

In this paper, we aim to investigate the possible usage and benefits of using the CS for the measurement and analysis of the von Kármán vortices. As indicated in Fig.~\ref{fig2}, due to the initialization adjustment of the numerical scheme, the initial part of the time series of  $C_D$ and $C_L$ do not exhibit a steady-in-the-mean behavior. Therefore, in order to avoid the numerical errors introduced in the initialization phase of the simulations, we allow for adjustment and analyze the time series of the $C_D$ within the temporal range of approximately $t \in [100s,200s]$.
In order to employ CS, it is necessary to investigate if the sparsity criteria are met. It is possible to employ CS if the time series has a sparse representation in a domain where orthogonal transformations such as the Fourier, cosine, or wavelet transforms are used for mapping from the physical space. Since the time series of the $C_D$ exhibits undulations about a mean level for this simulation, it is straightforward to see that it has a sparse representation in the frequency domain. In order to illustrate this behavior, we depict the two sided Fourier spectrum of the time history of the $C_D$ in Fig.~\ref{fig3}.
The first subplot depicted in Fig.~\ref{fig3} shows the Fourier spectrum of the time series of $C_D$ obtained using an FFT operation using $N=994$ classical samples. If longer time series are analyzed, the numbers of spectral components can be selected as a power of 2 for the efficient computations of FFT operations. Checking the first subplot of Fig.~\ref{fig3}, it is possible to realize that energy is mainly located at a central frequency due to the mean drag force. Minor energy exists at higher frequency values of the spectrum and the shedding frequency in the flow direction is at $f_{fl} \approx 0.54 Hz$ for this simulation. Thus, it is possible to state that the time series of $C_D$ has a sparse representation in the Fourier spectral domain. Since the sparse representation of this signal exhibiting oscillations due to von Kármán vortices, we take $M$ random samples in the physical domain and apply the $l_1$ minimization of the CS algorithm to these samples. As depicted in the second subplot of Fig.~\ref{fig3}, the same spectrum can be reconstructed using only $M=128$ compressive samples. We observe the near-perfect reconstruction of the spectrum by using only $M=128$ compressive samples. The term 'overwhelmingly high probability' is coined by the relevant literature on CS \cite{candes, candes2006compressive, candesRomberg2, candesTao, Baraniuk} to indicate the degree of reconstruction. By further increasing the number of compressive samples $M$, where still a large value of under-sampling ratio $r=N/M$ is satisfied, it is possible to obtain an exact reconstruction depending on the degree of sparsity, $K$, of the time series. Using IFFT operations, we reconstruct the time series of the $C_D$ from the spectrum depicted in the second subplot of Fig.~\ref{fig3} and we display the recovered time series in Fig.~\ref{fig4}.

\begin{figure}[h!]
\begin{center}
   \includegraphics[width=6.0in]{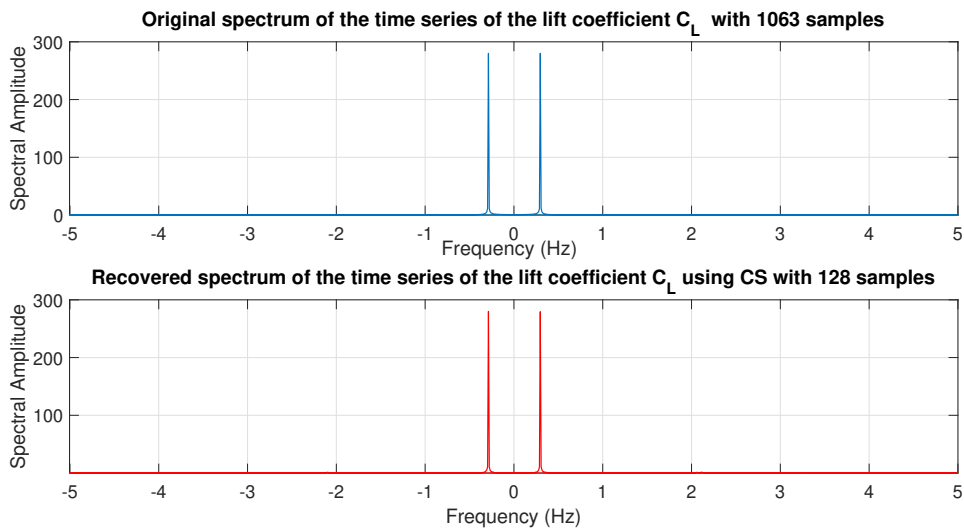}
  \end{center}
\caption{\small  Spectrum of the time series of the lift coefficient, $C_L$ of the flow around the circle: a)  classically sampled using $N=1063$ samples b) reconstructed using compressive sampling using $M=128$ samples.}
  \label{fig5}
\end{figure}

Next, we turn our attention to the reconstruction of the time series of the lift coefficient, $C_L$, imposed by the flow on the circle. As before, in order to avoid the numerical errors emerging due to the initialization of the numerical algorithm, we investigate the steady-in-the-mean part of the lift coefficient by using the temporal range of approximately $t \in [90s,200s]$. The two sided Fourier spectrum of the time series of the $C_L$ for this temporal range is depicted in Fig.~\ref{fig5}. Contrary to the time series of the drag coefficient, since the time series of the $C_L$ has a mean value of 0 for this temporal range, no energy is located at the frequency of $f=0$.

The first subplot in Fig.~\ref{fig5} shows the spectrum obtained by using $N=1063$ classical samples. Similar to the time series of the drag coefficient imposed on the circular cylinder by flow, the lift coefficient has also a sparse representation in the Fourier space as displayed. The shedding frequency in the transverse direction of the flow appears to be $f_{tr}\approx 0.27 Hz$. Thus, the Strouhal number for this flow becomes $St=f_{tr}D/U \approx 0.27 Hz \times 1m / 1m/s \approx 0.27$. Since the spectrum of the time history of the $C_L$ imposed on the circle is sparse, by taking random measurements in the time domain and using the $l_1$ minimization of the compressive sampling, it is possible to reconstruct this sparse spectrum using CS. In the second subplot of Fig.~\ref{fig5}, the same spectrum reconstructed by using $M=128$ random compressive samples is displayed.

\begin{figure}[h!]
\begin{center}
   \includegraphics[width=6.0in]{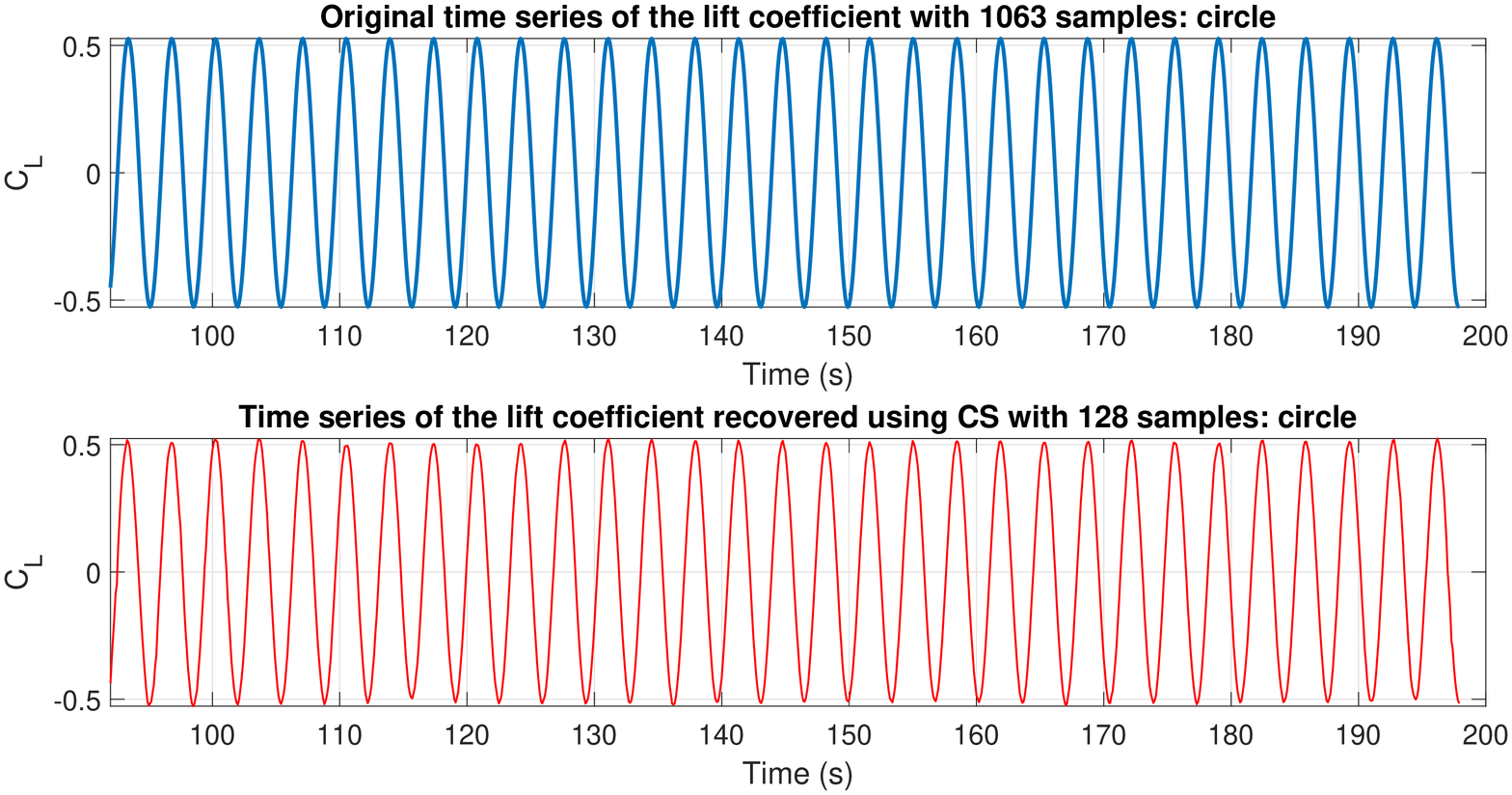}
  \end{center}
\caption{\small Time series of the lift coefficient, $C_L$, of the flow around the circle: a) classically sampled using $N=1063$ b) reconstructed using $M=128$ compressive samples.}
  \label{fig6}
\end{figure}
The time series of the lift coefficient, $C_L$, obtained by inverting the Fourier spectra depicted in Fig.~\ref{fig5} is displayed in Fig.~\ref{fig6}. As one can observe from the figure, the compressive sensing can be effectively used to reconstruct the time series of the lift coefficient by using a far fewer number of samples than its classical sampling analog.

\subsection{Results for the Flow Field Around An Ellipse}
\begin{figure}[h!]
\begin{center}
   \includegraphics[width=5.0in]{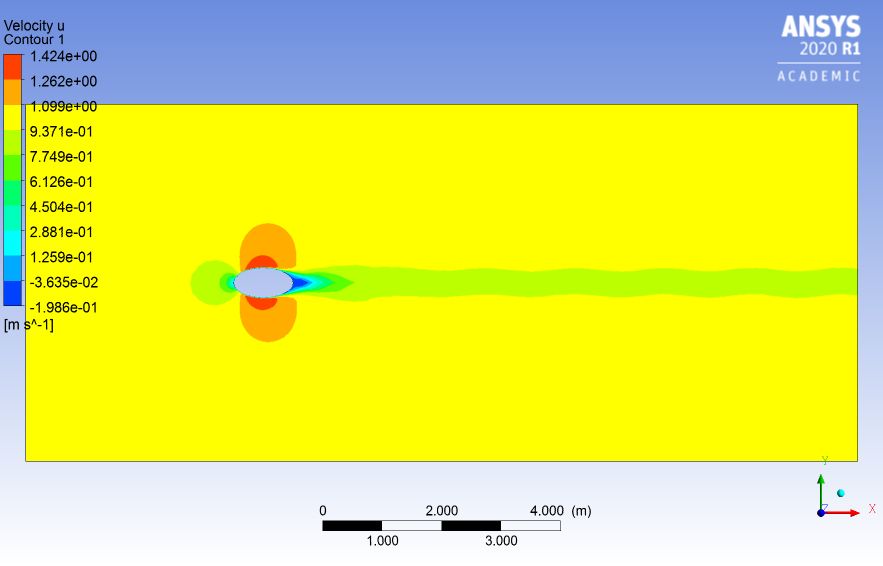}
  \end{center}
\caption{\small von Kármán vortices around an ellipse for $v=1m/s$ at $t=200s$. }
  \label{fig7}
\end{figure}

It is well-known that various blunt-body shapes can have a significant effect on the characteristics of the von Kármán vortices. Many possible designs that can reduce the effects of von Kármán vortices include but are not limited to helical strakes wrapped around circles and rounded, chamfered, sawtooth, and double notch corners around the square and rectangular cross-sections, just to name a few. It is well-known that blunt-bodies that have elliptical cross-sections can also be used to suppress these vortices \cite{MunsonYoungOkiishi, KunduCohen}. With this motivation, we investigate the usage of CS to effectively measure and analyze 2D von Kármán vortices around an elliptical cylinder. We consider the flow field around an ellipse having a width of $D_1=1m$ and a height of $D_2=0.5m$. The computation parameters are selected as $\rho=1.225 kg/m^3$ for the density of the air and $\mu=1.7894 \times 10^{-5} kg/(m.s)$ for its viscosity. As before, the Reynolds number for this flow becomes $Re=\rho U D/ \mu \approx 6.85 \times 10^{4}$, and the turbulence model is selected to be the $k-\epsilon$ model. For the temporal evolution of the numerical scheme, a time step value of $dt=0.1s$ is used until the time of $t=200s$. The contour map of the flow field around this elliptical cylinder exhibiting suppressed von Kármán vortices is depicted in Fig.~\ref{fig7}. 

\begin{figure}[h!]
\begin{center}
   \includegraphics[width=6.0in]{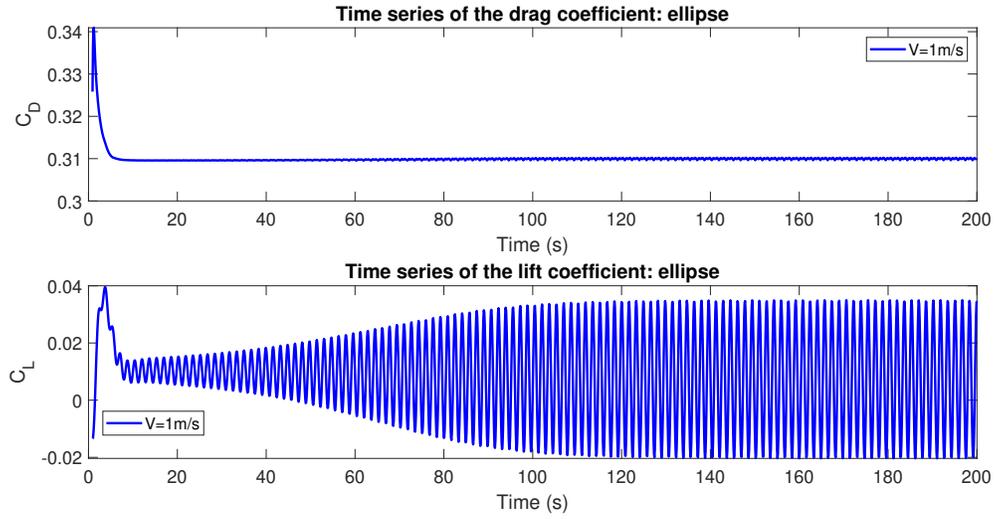}
  \end{center}
\caption{\small  Time histories of the von Kármán vortices around an ellipse for $v=1m/s$ a) Drag coefficient b) Lift coefficient.}
  \label{fig8}
\end{figure}

For this flow field, the time series of the drag and lift coefficients are displayed in Fig.~\ref{fig8}. As before, forthe initialization of the numerical scheme we allow for adjustment and analyze the time series of the $C_D$ and $C_L$  within the temporal range of approximately $t \in [120s,200s]$. Checking this figure, it is possible to observe the mean values of the steady-in-the-mean coefficients are $C_D \approx 0.3100$ and $C_L \approx 0.0005$. Additionally, one can observe that the maximum values of steady-in-the-mean coefficients are $C_D \approx 0.3105$ and $C_L \approx 0.0350$, which are in accordance with the well-established results tabulated in \cite{MunsonYoungOkiishi, KunduCohen}. Although the oscillations in the time series of $C_D$ and $C_L$ can be clearly observed due to the existence of the von Kármán vortices in the flow field around the ellipse, they are significantly suppressed compared to their analogs occurring in the flow field around the circle, as expected.

\begin{figure}[h!]
\begin{center}
   \includegraphics[width=6.0in]{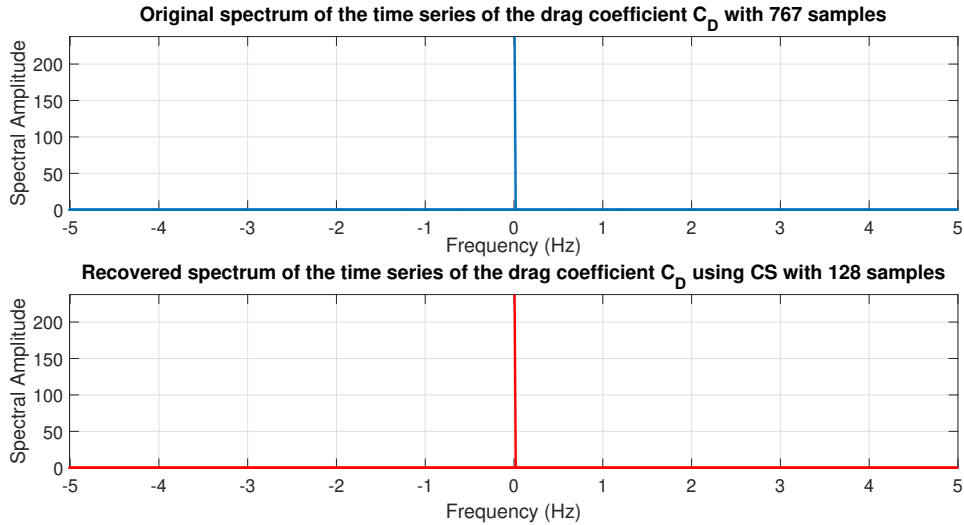}
  \end{center}
\caption{\small Spectrum of the time series of the drag coefficient, $C_D$ of the flow around the ellipse: a)  classically sampled using $N=767$ samples b) reconstructed using compressive sampling using $M=128$ samples.}
  \label{fig9}
\end{figure}
\begin{figure}[h!]
\begin{center}
   \includegraphics[width=6.0in]{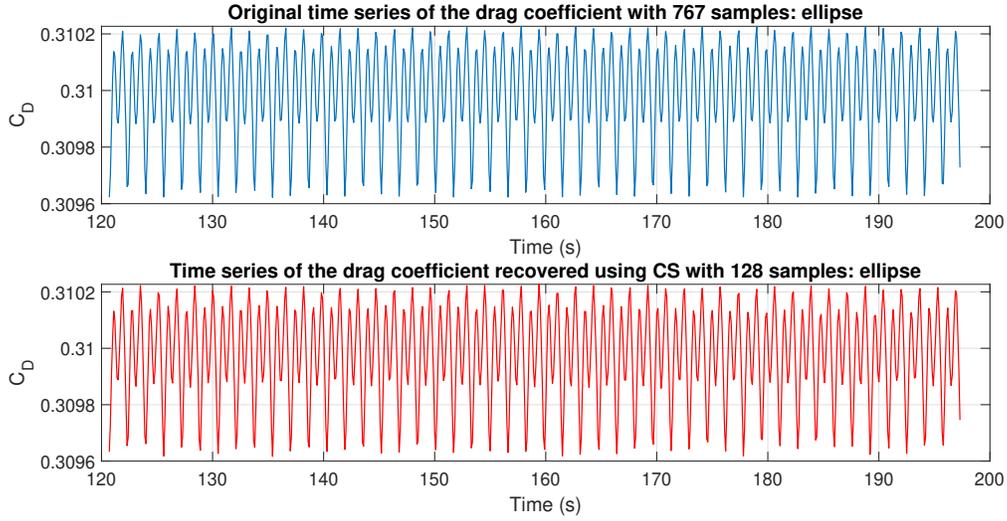}
  \end{center}
\caption{\small Time series of the drag coefficient, $C_D$, of the flow around the ellipse: a) classically sampled using $N=994$ b) reconstructed using $M=767$ compressive samples.}
  \label{fig10}
\end{figure}

In order to assess the applicability of the CS, we transform the time series depicted in Fig.~\ref{fig8} into the frequency domain by utilizing FFTs and plot the corresponding spectra in Fig.~\ref{fig9} for $N=767$ classical samples. Checking the first subplot in Fig.~\ref{fig9}, one can observe that time series of the $C_D$ and $C_L$ exerted by the flow of air onto ellipse have sparse representations in the Fourier space, thus CS can used for their efficient reconstruction. The spectrum depicted in first subplot of Fig.~\ref{fig9} exhibits significant energy located at the central $f=0$ frequency due to mean of value of the drag coefficient, however only very minor energy is located at higher frequency values of the spectrum. The two dominant shedding frequencies in the flow direction is found to be $f_{fl_1} \approx 0.62 Hz$ and $f_{fl_1} \approx 1.22 Hz$. Using $M=128$ random compressive samples and the $l_1$ minimization of the CS it is possible to reconstruct the spectrum as depicted in the second subplot of Fig.~\ref{fig9}.
After the construction of this spectrum, it is also possible to reconstruct the time series using IFFT operations. Classically sampled original time series of the drag coefficient with $N=767$ samples and its reconstruction by $M=128$ compressive samples are displayed in Fig.~\ref{fig10} for comparison purposes.

\begin{figure}[h!]
\begin{center}
   \includegraphics[width=6.0in]{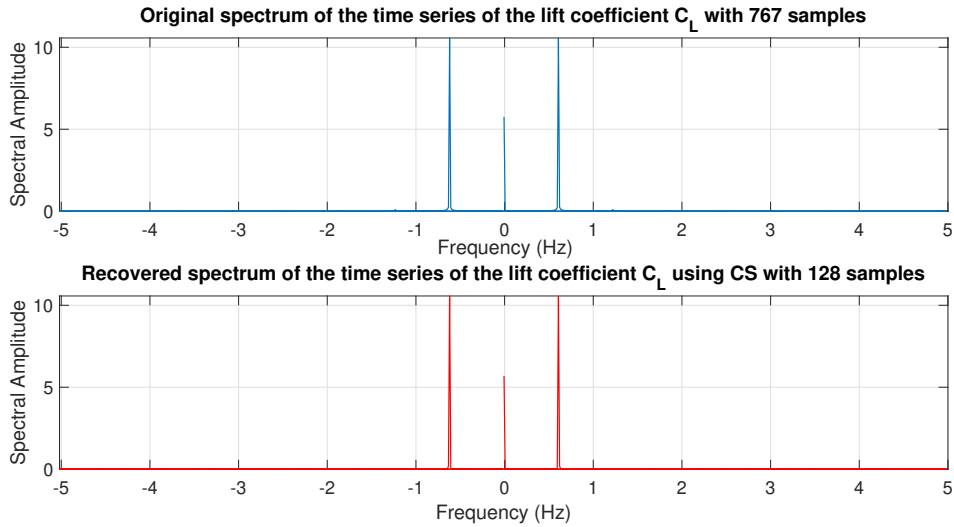}
  \end{center}
\caption{\small Spectrum of the time series of the lift coefficient, $C_L$ of the flow around the ellipse: a)  classically sampled using $N=767$ samples b) reconstructed using compressive sampling using $M=128$ samples.}
  \label{fig11}
\end{figure}

\begin{figure}[h!]
\begin{center}
   \includegraphics[width=6.0in]{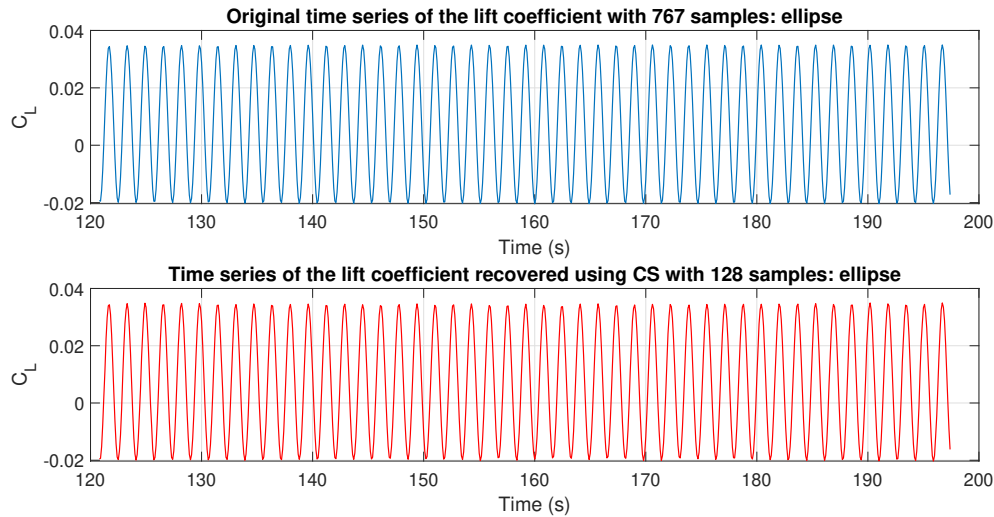}
  \end{center}
\caption{\small Time series of the lift coefficient, $C_L$, of the flow around the ellipse: a) classically sampled using $N=767$ b) reconstructed using $M=128$ compressive samples.}
  \label{fig12}
\end{figure}

After carrying out a similar analysis for the reconstruction of the spectrum and the time history of the lift coefficient around the elliptical cylinder, we depict the resulting spectra in Fig.~\ref{fig11} and reconstructed time series of the $C_L$ in Fig.~\ref{fig12}.

As before, the number of classical samples is $N=767$ and the number of classical samples is $M=128$. The spectrum depicted in the first subplot of Fig.~\ref{fig11} exhibits energy located at the central frequency of $f=0$ due to the mean value of the $C_L$, as well as significant energy located at the transverse oscillation frequency of $f_{tr}=60 Hz$. Thus, the Strouhal number for this flow fields around the elliptical cylinder becomes $St=f_{tr}D/U \approx 0.60 Hz \times 1m / 1m/s \approx 0.60 $ where $f_{tr}$ denotes the shedding frequency in the transverse direction to the flow. By employing IFFTs, it is possible to reconstruct the time series of the $C_L$ as depicted in Fig.~\ref{fig12}. 

The advantage of using CS for the measurement, analysis, and reconstruction of the time series of the von Kármán vortices is obvious. Depending on the number of samples $N$, the sparsity of the time series $K$ and the number of compressive samples $M$, the exact recovery of these time series are possible for $M<<N$. Since CS requires a significantly smaller number of samples for the exact reconstruction of the sparse signals, time series of the von Kármán vortices and similar phenomena can be efficiently measured and analyzed by the CS. This brings the advantage of reduced memory requirement, reduced sensor deployment, and fewer measurements. These advantages also enhance the cost as well as the processing times which can be critically important for the real-time measurement and vibration control of the engineering structures. Also, CS can also be used for the extrapolation and interpolation of the von Kármán vortices and similar flow phenomena using an approach similar to the one discussed in \cite{Malara}.

The CS-based approach presented in this paper for the efficient measurement and analysis of the time series of $C_D$ and $C_L$ can be extended to the measurement and analysis of different flow parameters. Two of the most important ones of these parameters are the drag force, $F_D$ and the lift force, $F_L$ given by
\begin{equation}
\begin{split}
& F_D= \frac{1}{2}C_D \rho U |U|A_D  \\
& F_L= \frac{1}{2}C_L \rho U |U|A_L
\label{eq11}
\end{split}
\end{equation}
where $A_D$ is the area perpendicular to the fluid flow and $A_L$ is the planform area. Such efficient measurements would lead to efficient and faster computations of the fluid forces and structural response. Structural vibration control can also be performed by feedback forces that are efficiently measured and analyzed by CS.

There are many possible directions for the generalization of our work. First of all, it is possible to extend our work to 3D. Additionally, the von Kármán vortices and similar phenomena occurring in the flow with multiple bluff-bodies can also be investigated and is very likely to give positive results. It is useful to remember that different time series of the flow parameters can have a sparse representation in the temporal or spatial domain, or any other transformed domain mapped using orthogonal transformations line DCT, wavelet, curvelet, ridgelet or chirplet transforms. Therefore, for different flow types, geometries, operating conditions, and engineering structures such as the morphing aircraft, moored offshore platforms and supersonic fighter jets launching missiles, obtaining the sparse representation of the flow parameters and their reconstruction by CS remains a future area of research.

\section{Conclusion}

In this paper, we investigated the possible usage and the implementation of the compressive sensing for the efficient measurement and analysis of the von Kármán vortices and similar phenomena occurring in fluid flows around blunt bodies. With this motivation, we numerically simulated the flow fields around two types of blunt bodies, a circular cylinder and an elliptical cylinder. By solving the governing $k-\epsilon$ transport equations, we modeled the time histories of the drag and the lift coefficients. We have shown that these time series have a sparse representation in the spectral domain, thus compressive sensing can be effectively used as an efficient tool for the measurement and the analysis of such time series of the von Kármán vortex parameters. The main advantage of the compressive sensing in measuring and analyzing such phenomena is its ability to allow for exact reconstruction of these time series by using a far fewer number of measurements compared to its classical analog of Shannon's theorem states. In potential applications the compressive sensing algorithm can be used for the development of efficient sensors and hardware, as well as a data analysis tool to extrapolate and interpolate the von Kármán vortex data. Dynamics of coastal and offshore structures, bridges, tall structures, airplanes and Bose-Einstein condensation are a few of the areas for which our findings will become very beneficial. It is also possible to extend our results to higher dimensions, other types of flows involving multi-blunt bodies having sparse representations in various orthogonal domains such as time, wavelet, ridgelet, or chirplet domains.

\end{document}